\documentclass[conference]{IEEEtran}

\usepackage{cite}
\usepackage{amsmath,amssymb,amsfonts}
\usepackage{graphicx}
\usepackage{textcomp}
\usepackage[dvipsnames]{xcolor}
\usepackage{array}
\usepackage{booktabs}
\usepackage{paralist}
\usepackage{xspace}
\usepackage{balance}

\ifCLASSOPTIONcompsoc
    \usepackage[caption=false,font=normalsize,labelfont=sf,textfont=sf]{subfig}
\else
    \usepackage[caption=false,font=footnotesize]{subfig}
\fi

\usepackage{todonotes}

\def\BibTeX{{\rm B\kern-.05em{\sc i\kern-.025em b}\kern-.08em
    T\kern-.1667em\lower.7ex\hbox{E}\kern-.125emX}}

\newcommand{\pros}{\textcolor{ForestGreen}{$\oplus$}\xspace}
\newcommand{\cons}{\textcolor{Red}{$\ominus$}\xspace} 

\begin{document} 

\title{Towards debiasing code review support}


\author{\IEEEauthorblockN{Tobias Jetzen}
    \IEEEauthorblockA{\textit{NADI},
    \textit{University of Namur}\\
    Namur, Belgium}
    \and
    \IEEEauthorblockN{Xavier Devroey}
    \IEEEauthorblockA{\textit{NADI},
    \textit{University of Namur}\\
    Namur, Belgium \\
    xavier.devroey@unamur.be}
    \and
    \IEEEauthorblockN{Nicolas Matton}
    \IEEEauthorblockA{\textit{NADI},
    \textit{University of Namur}\\
    Namur, Belgium \\
    nicolas.matton@unamur.be}
    \and
    \IEEEauthorblockN{Beno\^it Vanderose}
    \IEEEauthorblockA{\textit{NADI},
    \textit{University of Namur}\\
    Namur, Belgium \\
    benoit.vanderose@unamur.be}
}

\maketitle

\begin{abstract}
Cognitive biases appear during code review. They significantly impact the creation of feedback and how it is interpreted by developers. These biases can lead to illogical reasoning and decision-making, violating one of the main hypotheses supporting code review: developers' accurate and objective code evaluation. This paper explores harmful cases caused by cognitive biases during code review and potential solutions to avoid such cases or mitigate their effects. In particular, we design several prototypes covering confirmation bias and decision fatigue. We rely on a developer-centered design approach by conducting usability tests and validating the prototype with a user experience questionnaire (UEQ) and participants’ feedback. We show that some techniques could be implemented in existing code review tools as they are well accepted by reviewers and help prevent behavior detrimental to code review. This work provides a solid first approach to treating cognitive bias in code review.
\end{abstract}

\begin{IEEEkeywords}
cognitive bias, code review, user-centered design\vspace{-5mm}
\end{IEEEkeywords}


\section{Introduction}
\label{sec:introduction}

One of the many software development activities taken to ensure code quality is \textit{code review}. Code review consists of methodical code assessments that follow pre-defined guidelines and are supported by various tools to identify potential bugs, increase code readability and understandability, help developers learn the source code (i.e., code knowledge transfer), etc. Practically, code review is performed by a developer (i.e., a \textit{reviewer}), usually other than the \textit{author} of the code being reviewed. This reviewer, potentially helped by various tools, will read the code and make comments, ask questions, and request changes in the code that the author will take care of. Once the author and the reviewer are satisfied, the code is included (i.e., \textit{merged}) into the code base. 
The main goals of code review are to prevent defects, enable knowledge, check the code readability, enforce maintainability standards, etc., on recently modified source code \cite{badampudi_modern_2023}. Code reviews also serve as gatekeepers to prevent developers from committing arbitrary code without verification \cite{sadowski_modern_2018}. 

Like many software engineering activities, code review not only involves applying technical knowledge but heavily relies on social interactions between the reviewer and the author of the code \cite{mohanani_cognitive_2020, fagerholm_cognition_2022, chattopadhyay_tale_2020, salman_controlled_2019}. Such social-based activities are heavily influenced by cognitive and social aspects often neglected. 
This research focuses on \textit{cognitive biases} \cite{arkes_costs_1991}, and more specifically, their triggers and potential effects on code review quality. Our first goal is to identify factors triggering cognitive biases for the reviewer or the author. Our second goal is to design solutions to avoid such biases or mitigate their effects.

In this short paper, we design solutions addressing the triggers of \textit{confirmation bias} and the effects of \textit{decision fatigue}. For that, we follow a user-centered (here, developer-centered) design approach \cite{mao2005state, norman1986user, norman2013design}. In the first phase, we explore potentially harmful situations and design theoretical solutions to prevent or mitigate biases. In the second phase, we aim to improve the designed solutions by conducting usability tests with beginners using a prototype based on an existing code review tool. These tests serve as feedback to gather the users' requirements for an acceptable solution. To achieve this, we iterate multiple times over the prototype. Finally, we conduct an evaluation of the prototype’s final result by testing the user experience with the standard \textit{User Experience Questionnaire} (UEQ) \cite{schrepp_construction_2017, hutchison_applying_2014}. With the prototype as a final result and an evaluation of its usability (see our replication package \cite{replicationpackage}), we propose a first work on solving problematic relationships between cognitive bias and code review. The prototype will serve as the basis for developing a functional tool to evaluate the impact of our solutions on confirmation bias and decision fatigue.

\section{Background and related work}
\label{sec:background}




In psychology, cognitive biases (denoted biases hereafter) refer to instances where human cognition consistently generates representations that are systematically distorted when compared to objective reality \cite{haselton_evolution_2015}. Unlike logical fallacies, which are arguments based on invalid reasoning, biases are patterns of thinking that affect how we interpret new information and processes. Biases are applied systematically and influence our behavior, opinions, and decisions. 
Causes of biases are rooted in heuristics \cite{kai1979prospect, haselton_evolution_2015, tversky_judgment_1974, arkes_costs_1991, kahneman_thinking_2011}: shortcuts or rules of thumb used by our brain to solve a problem or judge a situation quickly. For instance, people with a higher social position tend to apply stereotypical views on others more often than those with a more precarious position, who invest more time and energy in social judgment \cite{haselton_evolution_2015}. However, in general, precisely identifying the exact causes of a specific bias is challenging \cite{mohanani_cognitive_2020}.

To eliminate biases (i.e., \textit{debiasing}), previous research has shown that neither applying more effort nor being more experienced in a field helps mitigate cognitive biases \cite{fischoff_debiasing_1981}. However, training on cognitive biases and applying specific techniques can make a substantial difference. This has been proven not only for experts in a field but also to affect the judgment of non-experts \cite{da2015towards}. 

\begin{figure}[t]
    \centering
    \includegraphics[width=0.44\textwidth]{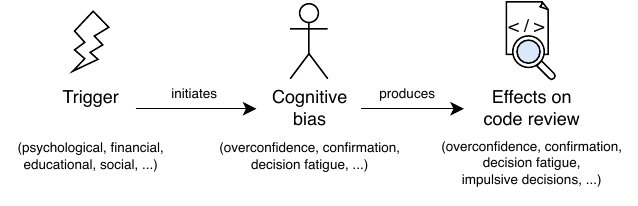}
    \caption{Relationship between triggers, cognitive biases, and their effects.}
    \label{fig:cognitive-bias}
    \vspace{-6mm}
\end{figure}

Practically, as illustrated in Figure \ref{fig:cognitive-bias}, when we know a cognitive bias, we can identify \textbf{triggers} initiating it and what \textbf{impact} the bias has on the investigated activity. In this work, a trigger is considered a specific environmental condition, enabling a cognitive bias. One cognitive bias can potentially be triggered by multiple elements \cite{ralph2011introducing}. Once triggered, it affects the person’s activity. In our case, code reviews can be impacted severely by cognitive biases. One cognitive bias, in turn, can produce multiple effects during the different activities.

Software engineering being a process heavily relying on human effort and involvement, cognitive biases have also been explored for various software engineering activities \cite{mohanani_cognitive_2020, fagerholm_cognition_2022, fleischmann_cognitive_2014, chattopadhyay_tale_2020, salman_controlled_2019}.
For instance, Barroso et al. \cite{barroso_influence_2017} investigated developers' personalities' influence on their tasks. They have shown that the quality of the product depends on the interaction between members of the team combined with their professional capabilities and, therefore, on the interpretation and reactions to feedback.  
Going further, Spadini et al. \cite{spadini_primers_2020} investigated the effect of existing review comments on code review and showed that reviewers are subject to availability bias when performing reviews.
More recent research has evidenced the importance of appropriate techniques to deal with cognitive biases during code review, such as checklists to potentially lower developers’ cognitive load \cite{goncalves_explicit_2022}; avoiding destructive criticisms not to decrease motivation \cite{gunawardena_destructive_2022}; or guideline to deal with confusion during code reviews \cite{ebert_exploratory_2021}.

In a recent study, Fagerholm et al. \cite{fagerholm_cognition_2022} have identified several future research directions for cognition in software engineering, including perception and software quality, which have still received very little attention. In this work, we focus on debiasing code review to avoid confirmation bias (i.e., \textit{triggers}) and mitigating potential \textit{effects} of decision fatigue. 


\paragraph{Confirmation bias}
One of the most researched cognitive biases in psychology is the confirmation bias. When talking about confirmation bias, one refers to the collection, interpretation, analysis and research for information in a way that confirms one’s prior beliefs instead of searching for information disproving them \cite{jorgensen_believing_2015, rainer_follow-up_2008}. In practice, once the mind adopts an opinion, it does everything to support it, leading to wrong decisions defying the sense of logical reasoning. For instance, the positive test bias leads developers to test only to confirm the code, instead of disproving it \cite{teasley_why_1994, salman_controlled_2019}. Tests are more effective with data, which is designed to disconfirm hypothesis \cite{leventhal_analyses_1994}. In general, and not only during tests, one’s goal should be to fail the code in order to reduce defect density \cite{calikli_empirical_2010}.

\paragraph{Decision fatigue}
A high number of decisions to make, each requiring to process information, over a short period of time leads to the depletion of internal resources, also referred to as \textit{ego depletion} \cite{baumeister_ego_1998}. When ego depletion manifests as decision fatigue, it causes attentional deficit, impulsive decisions, and leads to postponing decisions with the intention to look at them later \cite{danziger_extraneous_2011}. Finally, people subjected to decision fatigue tend to have an impaired ability to make trade-offs; they prefer acting in a passive role and make irrational judgments. Unfortunately, such changes in behavior are hard to recognize \cite{pignatiello_decision_2020}.

\section{Debiasing code review}
\label{sec:approach}

Our approach addresses the triggers and effects of confirmation bias and decision fatigue. As mentioned in Section \ref{sec:background}, precisely identifying the exact causes and effects of a specific bias is challenging. In our first step, we define scenarios in which triggers and effects can clearly be identified and reproduced. Exploring more complex triggers and effects of different biases is part of our future work.

\begin{table}[t]
    \centering
    \caption{Scenarios for Confirmation bias: triggers, effects, and remedies}
    \label{tab:confirmationbias}
    \begin{scriptsize}
    \begin{tabular}{r p{68mm}}
    \toprule
        \textbf{Trigger:}   & The developer gets low-quality feedback, hurting their self-esteem. \\
        \textbf{Effect:}   & The developer refuses recommendations from the feedback to protect their self-esteem.\\
        \textbf{Remedies:}  & \textit{Constructive feedback}. Prevent the bias by providing the reviewer with advice about how to give constructive feedback.\\ 
            & \textit{Review feedback}. Prevent the bias by suggesting to the reviewer to ask another developer for feedback about their review.\\ 
            
        \midrule
        \textbf{Trigger:}   & The reviewer is under time-pressure due to circumstances.\\
        \textbf{Effect:}    & The reviewer tries to validate the existing code instead of analyzing it objectively.\\
        \textbf{Remedy:}    & \textit{Encourage brainstorming}. Mitigate the impact by providing a form with empty solution fields to encourage the reviewer to think about multiple solutions.\\
    \bottomrule
    \end{tabular}
    \end{scriptsize}
    \vspace{-5mm}
\end{table}

\paragraph{Confirmation bias}
Research in psychology investigated confirmation bias a lot \cite{klayman_varieties_1995, calikli_empirical_2010, salman_controlled_2019}, providing a solid basis to research its relations to software engineering, especially to modern code review. We focus on two scenarios described in Table \ref{tab:confirmationbias}, with potential solutions. The assumption is that how a reviewer builds the feedback influences the developer’s perception and, therefore, their acceptance of the feedback (first line in Table \ref{tab:confirmationbias}). Also, when reviewers see code changes, they are exposed to code that influences their perception during review \cite{spadini_primers_2020}. Our assumption is that under time pressure, this phenomenon becomes more pronounced: a reviewer tends to search for fast review approval instead of correct implementation (second line in Table \ref{tab:confirmationbias}).

\paragraph{Decision fatigue}
Many triggers can initiate decision fatigue \cite{sievertsen_cognitive_2016, baumeister_ego_1998}. We focus on the scenarios described in Table \ref{tab:decisionfatigue}. 
The first and second scenarios consider that a reviewer needs motivation to tackle new code or potentially new topics. Humans tend to perform small tasks where they are rewarded early: this behavior is called hyperbolic discounting \cite{laibson_golden_1997}. However, during code review, a reviewer may get assigned a great number of reviews to do or review code requiring specific knowledge. Our assumption is that when decision fatigue is triggered due to circumstances that are unfavorable for starting tasks intensive in cognitive resources, the reviewer tends to procrastinate.
In the third scenario, decision fatigue leads the reviewer to make impulsive comments. Our assumption is that the comments will be expressed in a familiar way, leading to destructive feedback instead of constructive one.
In the last scenario, our assumption is that understanding the code is key to making constructive comments for the author. A review under the effect of decision fatigue (i.e., not taking all the elements into account) might provoke misleading results in the feedback for the author. 

Mitigating the impact of biases on code review requires dealing with either the trigger initiating the bias or its effects. Depending on the scenario, the former or the latter might be better suited.

\begin{table}[t]
    \centering
    \caption{Scenarios for Decision fatigue: triggers, effects, and remedies}
    \label{tab:decisionfatigue}
    \begin{scriptsize}
    \begin{tabular}{r p{68mm}}
    \toprule
        \textbf{Trigger:}   & The reviewer is over-solicited.\\
        \textbf{Effect:}    & The reviewer misses motivation to do reviews and postpones them for later (i.e., procrastination).\\
        \textbf{Remedies:}  & \textit{Scheduled reviews}. Prevent the bias by limiting the amount of reviews to a maximum number and a calendar to schedule.\\ 
            & \textit{Observe needed time}. Prevent the bias by reminding the reviewer to halt when too much time is needed for review.\\  
        
        \midrule
        \textbf{Trigger:}   & The reviewer misses knowledge about a specific topic in the code.\\ 
        \textbf{Effect:}    & The reviewer misses motivation to do reviews and postpones them for later (i.e., procrastination).\\
        \textbf{Remedy:}    & \textit{Find an expert}. Prevent the bias by assigning the best fitting reviewer according to their experience in the topic.\\ 
        
        \midrule
        \textbf{Trigger:}   & The reviewer is working at times of day known for decreased internal resources (e.g., the end of the working day or after lunch).\\
        \textbf{Effect:}   & The reviewer makes impulsive comments instead of constructive suggestions for the author.\\
        \textbf{Remedy:}    & \textit{Guide with comments}. Mitigate the impact by guiding the reviewer through the files with comments made by the author.\\ 
        
        \midrule
        \textbf{Trigger:}   & The reviewer lacks of experience in making code reviews.\\
        \textbf{Effect:}   & The reviewer skips code changes or parts of the code, leading to a lower understanding of the code.\\
        \textbf{Remedy:}    & \textit{Help commenting}. Mitigate the impact by providing a form with keywords to help the reviewer to include all essential elements.\\ 
    \bottomrule
    \end{tabular}
    \end{scriptsize}
    \vspace{-2mm}
\end{table}

\section{Design of the code review support}
\label{sec:implementation}

\begin{figure}[t]
    \centering
    \includegraphics[width=0.38\textwidth]{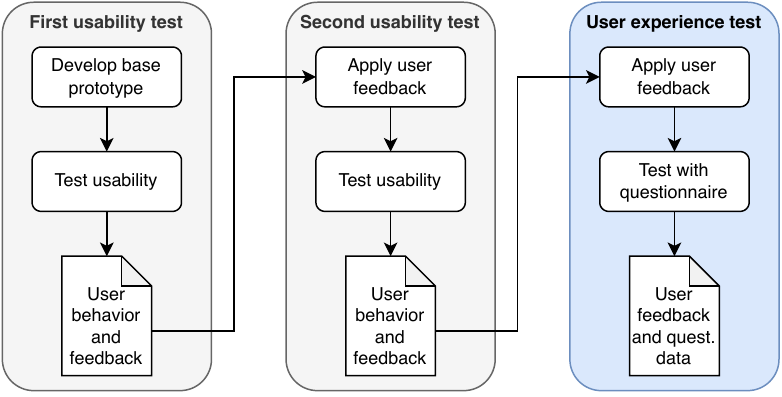}
    \caption{Design, prototyping, and evaluation}
    \label{fig:methodology}
    \vspace{-5mm}
\end{figure}

In the following section, we will design and test potential solutions to address the remedies described in Tables \ref{tab:confirmationbias} and \ref{tab:decisionfatigue}. For that, we follow a user-centered (here, reviewer/developer-centered) design process \cite{mao2005state, norman1986user, norman2013design} to guide the building of our support following an iterative approach.
Figure \ref{fig:methodology} provides an overview of the process: due to time constraints, we limited the development to two iterations (i.e., \textit{usability tests}), concluded by a \textit{user experience test} of the designed solutions, performed with a different group of users than the one involved in the first and second usability tests. 
The tested solutions are developed as HTML prototypes. We relied on two groups of users: three for the first and second usability tests and five for the final user experience test. We employ a total of eight participants, which, according to Faulkner \cite{faulkner_beyond_2003}, helps identify, on average, more than 80\% of the problems.

\subsection{First usability test}


First, we create an initial prototype based on the web tool Gerrit\footnote{https://www.gerritcodereview.com}, similar to what can be found on GitHub.
Then we implement the remedies listed in Tables \ref{tab:confirmationbias} and \ref{tab:decisionfatigue} using the following techniques. All the prototypes can be found in our replication package \cite{replicationpackage}.


\paragraph{Technique 1.1 –- Advice} 
The goal is to help the reviewer create constructive feedback \cite{gunawardena_destructive_2022}. The first technique is based on advice displayed in the form of a list. 
This list can be opened by clicking the button \textit{I need advice}. Following this action, a popup shows a list containing the advice. The advice originates from literature investigating how to achieve constructive feedback \cite{bee_constructive_1998, waggoner_denton_improving_2018}.


\paragraph{Technique 1.2 –- Form} 
Because under decision fatigue reviewers tend to make incomplete comments \cite{goncalves_explicit_2022}, the idea of a pre-structured form 
seems an appropriate choice. This way, the reviewer does not have to think about how to structure the comment. It should include the identification of a problem, a justification regarding why the discovery is considered a problem and also a suggestion to solve it. Here, three empty fields are available to encourage the reviewer to brainstorm multiple suggestions, thereby promoting more thorough and considered feedback.


\paragraph{Technique 1.3 -– Guide}
To avoid the user skipping changes or even entire files from being reviewed, a guide 
is offered just before starting the review \cite{ebert_exploratory_2021}. This guide consists of a certain amount of comments written by the author. They contain an explanation of why a certain change was made. When launching the guide, the reviewer’s attention is immediately drawn to the first comment, surrounded by a red border, as decided by the author. When the reviewer decides that they understand the change, they click on the button \textit{Next} to go to the next comment. Once having been through every step, the reviewer starts the actual review.

The test analyses the navigation through the prototype, usage of the tool, and reaction to the tool by the participants who act as developers or as reviewers, depending on the technique. Due to space constraints., we only report a summary of our observations in Table \ref{tab:usabilitytests}. Detailed protocol and results are available in our replication package \cite{replicationpackage}.
 
\begin{table}[t]
    \centering
    \caption{Summary of the usability tests observations}
    \label{tab:usabilitytests}
    \begin{scriptsize}
    \begin{tabular}{l p{60mm}}
    \toprule
        \multicolumn{2}{c}{Usability test 1}\\ 
        \midrule        
        Technique 1.1 & \pros Most users apply the advice after reading it.\\
        Advice & \pros The short formulated advice is appreciated.\\
            & \cons The popup button is not always noticed.\\
            & \cons Not everybody wants to be advised.\\
            & \cons Background color confuses participants.\\
        \midrule        
        Technique 1.2 & \pros All fields get filled out.\\
        Form & \pros The form provides a coherent structure.\\
            & \cons Only one solution is given.\\
            & \cons Some feel overwhelmed.\\
        \midrule
        Technique 1.3 & \pros Everyone uses the guide. \\
        Guide & \pros No code change is skipped.\\
            & \cons The guide could bias the reviewer’s comment.\\
            & \cons The \texttt{Next} button is not intuitive.\\
    \toprule
        \multicolumn{2}{c}{Usability test 2}\\
        \midrule 
        Technique 2.1 & \pros The advice is noticed and read immediately.\\
        Advice & \pros The advice impacts the overall comment.\\
            & \cons Green color signifies already complete.\\
            & \cons Participants mistake inciting items for to-do items.\\
        \midrule
        Technique 2.2 & \pros All participants use the technique.\\
        Example & \pros Saves time to think about structure.\\
            & \pros Comment analysis could be automated.\\ 
            & \cons Participants type the keywords manually.\\
        \midrule
        Technique 2.3 & \pros All participants use the technique.\\
        Quick search & \cons Some only use the quick search, without commenting.\\
        \midrule
        Technique 2.4 & \cons Most participants do not use the technique.\\
        Expert feedback & \cons Most users are concerned about annoying colleagues.\\
        \midrule
        Technique 2.5 & \pros Launch button is noticed faster.\\
        Guide & \pros Understanding the Next button is intuitive.\\
            & \pros The Next button acts as an obligation to comment.\\
            & \cons Unable to make comments inside the guide.\\
    \bottomrule
    \end{tabular}
    \end{scriptsize}
    \vspace{-5mm}
\end{table}

\subsection{Second usability test}

We improved the prototype following the observed behavior and the participants' suggestions. Some techniques get improved, two new techniques are added, and one is removed. 


\paragraph{Technique 2.1 –- Advice}
In the current iteration, the advice popup was transformed into a drop-down list that is immediately visible when the comment tool is opened. This change ensures that every participant notices and reads the advice, unlike in previous iterations. As stated in Table \ref{tab:usabilitytests}, the green background does not show positive effects. Also, though most elements can be used like in a checklist, some of them can not, because they are intended to incite the reviewer to analyze the code from another perspective.


\paragraph{Technique 2.2 –- Example}
In the previous iteration, a new technique emerged from the advice technique 1.1. Here, the technique uses an example, and participants follow the structure presented in the example. The feedback confirms appreciation, as shown in Table \ref{tab:usabilitytests}.


\paragraph{Technique 2.3 –- Quick search}
As requested in the first iteration, tasks should take less effort (i.e., fewer intermediary clicks). The form used before is now replaced with a quick search. It was used in different ways: first commenting, then searching for code snippets related to the comment, or searching for a solution before commenting. In the latter case, however, the comment only refers to the selected solution from the search without further explanation.


\paragraph{Technique 2.4 –- Expert feedback}
As for the previous technique, another help was requested: providing expert feedback allows for countering the effects of decision fatigue when the reviewer is prone to take a passive role. However, as observed during the tests, most participants do not use the expert feature because they feel they are experienced enough or do not want to annoy a senior colleague. 


\paragraph{Technique 2.5 –- Guide}
In this iteration, the launch button for the guide is bigger and thus better visible, and the button used to get to the next step is placed in a more understandable location. As stated in Table \ref{tab:usabilitytests}, it has a positive impact. However, all participants complain about not being able to give comments while using the guide.

\subsection{Final prototype}

\begin{figure}[!t]
    \centering
    \subfloat[Combination of advice and example.]{
        \includegraphics[width=60mm]{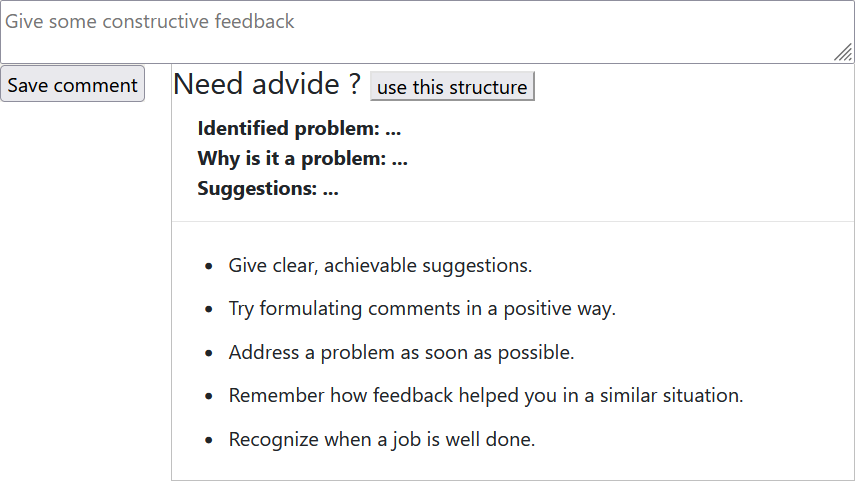}
        \label{fig:secondt1t2}
    }
    \hfil
    \subfloat[Combination of quick search and expert feedback.]{
        \includegraphics[width=70mm]{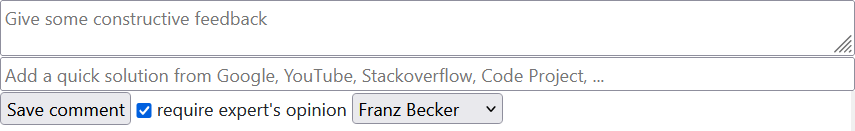}
        \label{fig:secondt3t4}
    }
    \caption{Combinations of techniques.} 
    \label{fig:secondcombination}
    \vspace{-5mm}
\end{figure}

During the second usability test, multiple participants suggested combining techniques. 

\paragraph{Technique 3.1 -- Advice}
Techniques 2.1 and 2.2 are combined to create constructive feedback, as displayed in Figure \ref{fig:secondt1t2}. A new button is added to allow preparing the comment structure faster: it copies the key phrases ``Identified problem'', ``Why is it a problem'' and ``Suggestions'' into the comment field. The advice is located beneath the example.

\paragraph{Technique 3.2 -- Assistant}
Techniques 2.3 and 2.4 are combined to mitigate decision fatigue's effects (Figure \ref{fig:secondt3t4}). 

\paragraph{Technique 3.3 -- Guide}
Addressing the negative point of Technique 2.5 (\cons in Table \ref{tab:usabilitytests}) would not support the intended effect. Being able to comment continuously hinders the reviewer from following the guide to understand the code as a whole and not only partially. Therefore, Technique 3.3 remains equivalent to Technique 2.5. 

\section{User experience test}
\label{sec:evaluation}



The goal of this preliminary evaluation is to gain feedback about the user interface and its usability, not about the psychological aspects. Those aspects are left for future work. 


An evaluation proceeds with each of the five participants, all experienced developers, familiar with code review, as follows: 
\begin{inparaenum}[(i)]
   \item we ask the participant about their experience in code review;
   \item depending on the answer, we explain and demonstrate what code reviews are and how tools can be used to assist the code review process;
   \item for each technique (Technique 3.1-3.3), we explain to the participant about the task (one per technique) to perform and proceed with the test. The explanation however takes place without mentioning the investigated aspects concerning cognitive biases. This is necessary to not bias the participant’s opinion.
   \item after completing each task, the participant is asked to fill out a standard User Experience Questionnaire (UEQ) \cite{schrepp_user_2019, hutchison_applying_2014, holzinger_construction_2008, schrepp_design_2017}.
\end{inparaenum} 
The UEQ is a heavily validated state-of-the-art questionnaire measuring user experience following the predefined scales: 
Attractiveness (do users like or dislike the product), 
Perspicuity (is it easy to get familiar with the product?), 
Efficiency (can users solve their tasks without unnecessary effort?), 
Dependability (does the user feel in control of the interaction?), 
Stimulation (is it exciting and motivating to use the product?),
and Novelty (is the product innovative and creative?). 
Additionally, we ask every participant after the questionnaire for personal feedback about the tested technique

\begin{figure}[!t]
    \centering
    \includegraphics[width=85mm]{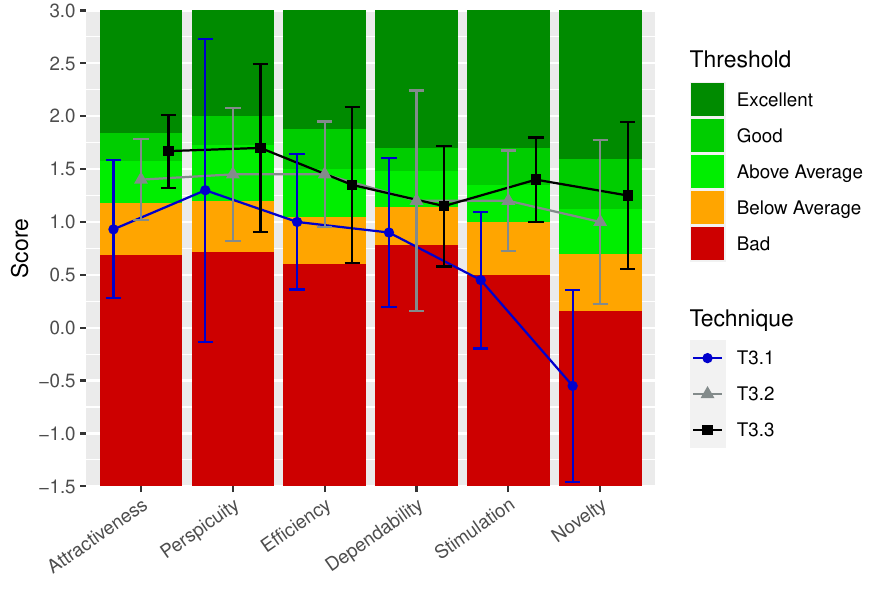}
    \caption{Results of the UEQ benchmark for the advice with example (T3.1), the quick search with expert feedback (T3.2), and the guide (T3.3). Dots denote mean values, and error bars indicate 95\% confidence intervals.} 
    \label{fig:ueqT3}
    \vspace{-5mm}
\end{figure}

We analyze questionnaire responses for each technique to evaluate if it improves user experience using the UEQ Data Analysis Tool \cite{schrepp_user_2019}. This tool provides a quantitative analysis, converting responses from a scale of 1 to 7 to a scale of -3 (most negative) to +3 (most positive). It classifies the scale values into five categories, from excellent to bad, based on a benchmark of user interfaces \cite{schrepp_construction_2017}. Results for the three techniques are reported in Figure \ref{fig:ueqT3}. Given the preliminary nature of the evaluation, the small sample size results in a less accurate quantitative analysis (as confirmed by the sometimes large confidence intervals in Figure \ref{fig:ueqT3}), yet it still indicates trends in user experience, supported by the additional qualitative feedback (not reported in this short paper due to space constraints). The complete data are available in our replication package \cite{replicationpackage}.


%
%
%
The UEQ results indicate that users perceive \textit{Technique 3.1 - Advice} as understandable and easy to learn. 
Users rated the technique poorly on the stimulation scale, indicating it was neither exciting nor motivating, and perceived it as conventional. 
%
%
%
%
Users also find \textit{Technique 3.2 - Assistant} interesting, exciting, and motivating, as well as efficient and easy to use, aligning with the technique's goals. The technique is seen as innovative in code reviews. 
Overall, the user experience results are positive, reflecting the feedback, but responses focus more on the \textit{quick search} feature than the \textit{expert feedback}.
Finally, users find \textit{Technique 3.3 - Guide} highly attractive and interesting, with clear indications of it being stimulating and innovative. It also receives positive evaluations for efficiency, with users feeling it meets expectations by providing support. 

\section{Discussion and future work}
\label{sec:discussion}

\subsection{Confirmation bias}

Our base assumption was that non-constructive feedback during code review provokes confirmation bias. 
Our preliminary results from applying \textit{Technique 3.1 - Advice} show that reviewers prefer guided, pre-defined structured comments but are generally unmotivated to follow written advice. Overall advice and examples are used quickly, without spending more than a few seconds to integrate them in the comments. 
Reviewers are willing to incorporate examples into the review process, with most agreeing on the positive effects of this technique. Feedback from our preliminary evaluation suggests that the technique helps prevent confirmation bias during code review.
While this research does not quantitatively measure the extent to which the technique prevents confirmation bias, it includes a prototype solution and tests user experience. Further investigation with a larger sample size is necessary for representative quantitative results.

\subsection{Decision fatigue}

\paragraph{Incomplete comments}
Our base hypothesis suggested that decision fatigue during code review leads to incomplete comments. From this, we proposed \textit{Technique 3.2 - Assistant}. 
User experience tests show that the search tool encourages adding code snippets to comments and is well-received once users understand its purpose. This technique shows significant potential to mitigate decision fatigue.
Conversely, the expert feedback tool is seen as annoying for senior colleagues, with reviewers avoiding it due to confidence in their comments. Currently, the expert feedback prototype cannot mitigate decision fatigue but could be improved through design adjustments.
Overall, the design and layout significantly influence the effectiveness of the tools. User understanding and willingness to use a technique depend heavily on the interface design, suggesting that alternative designs might yield better results.

\paragraph{Skipping code changes}
Our hypothesis also supports that decision fatigue during code review leads to skipping short, large, or complex changes. \textit{Technique 3.3 - Guide} can prevent this by helping reviewers address important changes individually. 
User experience tests show that reviewers consistently use the guide, following it to the end without skipping any changes. Participants appreciated the guide, indicating it effectively prevents decision fatigue from causing skipped reviews. Thus, the technique helps mitigate certain effects of decision fatigue during code review. 
%



\subsection{Future work}

The limited participant number constrained our quantitative data, and the high variance in some of the responses calls for further investigations. However, these results still helped identify tendencies and refine prototypes. 
Our future research will focus on design aspects for better outcomes, further prototype development, and extending the scope to other cognitive biases. User feedback also suggests automating review tasks and providing context-sensitive feedback. Ultimately, these prototypes could evolve into fully functional tools for real-world application.



\balance
\bibliographystyle{IEEEtranS}
\bibliography{references.bib}

\end{document}